%% file: main.tex
\documentclass[numbers]{article}
\usepackage{array}
\usepackage{ragged2e}
\usepackage[final]{neurips_2025}
\usepackage[utf8]{inputenc} 
\usepackage[T1]{fontenc}    
\usepackage{hyperref}       
\usepackage{url}            
\usepackage{pifont}         
\usepackage{url}            
\usepackage{booktabs}       
\usepackage{amsfonts}       
\usepackage{nicefrac}       
\usepackage{microtype}      
\usepackage{xcolor}         
\usepackage{booktabs}
\usepackage[table]{xcolor}
\usepackage{caption}
\usepackage{array} %
\usepackage{ragged2e} %
\usepackage[utf8]{inputenc}
\usepackage{amsthm}
\usepackage{algpseudocode}
\usepackage{tikz}
\usepackage{wrapfig}
\usepackage{amsmath}
\usepackage{multirow}
\usepackage{booktabs}
\usepackage{inconsolata}
\usepackage{adjustbox}
\usepackage{pifont}
\usepackage{enumitem}
\let\oldding\ding
\renewcommand{\ding}[2][1]{\scalebox{#1}{\oldding{#2}}}
\title{Mano Technical Report
}

\author{%
  Tianyu Fu
  \thanks{These authors contributed equally to this research} 
  \And
  Anyang Su
  $^*$ 
  \And
  Chenxu Zhao
  $^*$
  \thanks{Corresponding author.} 
  \And
  Hanning Wang
  $^*$ 
  \And
  Minghui Wu
  $^*$
  \thanks{Project leader.} \\
  \And
  Zhe Yu 
  \And
  Fei Hu
  \And
  Mingjia Shi 
  \And
  Wei Dong 
  \And
  Jiayao Wang 
  \And
  Yuyang Chen \\
  \And
  Ruiyang Yu 
  \And
  Siran Peng 
  \And
  Menglin Li 
  \And
  Nan Huang 
  \And
  Haitian Wei 
  \And
  Jiawei Yu \\
  \And
  Yi Xin \quad
  Xilin Zhao \quad 
  Kai Gu \quad
  Ping Jiang \quad 
   Sifan Zhou \quad
  Shuo Wang \\
  \\
  \\
  \textbf{DeepMiner-Mano Team, Mininglamp Technology}\\
  \footnotesize{\texttt{(futianyu, suanyang, zhaochenxu, wanghanning, wuminghui, wangshuo.e)@mininglamp.com}}
}

\begin{document}
\input{commands}
\newtheorem{theorem}{Theorem}
\newtheorem{lemma}{Lemma}
\newtheorem{remark}{Remark}
\newtheorem{corollary}{Corollary}
\newtheorem{assumption}{Assumption}
\newtheorem{proposition}{Proposition}
\newtheorem{definition}{Definition}

\maketitle
\newcommand{\say}[1]{\textcolor{blue}{#1}}
\begin{abstract}
Graphical user interfaces (GUIs) are the primary medium for human-computer interaction, yet automating GUI interactions remains challenging due to the complexity of visual elements, dynamic environments, and the need for multi-step reasoning. Existing methods based on vision-language models (VLMs) often suffer from limited resolution, domain mismatch, and insufficient sequential decision-making capability. To address these issues, we propose Mano, a robust GUI agent built upon a multi-modal foundation model pre-trained on extensive web and computer system data. Our approach integrates a novel simulated environment for high-fidelity data generation, a three-stage training pipeline (supervised fine-tuning, offline reinforcement learning, and online reinforcement learning), and a verification module for error recovery. Mano demonstrates state-of-the-art performance on multiple GUI benchmarks, including Mind2Web and OSWorld, achieving significant improvements in success rate and operational accuracy. Our work provides new insights into the effective integration of reinforcement learning with VLMs for practical GUI agent deployment, highlighting the importance of domain-specific data, iterative training, and holistic reward design.
\end{abstract}

\section{Introduction}

\input{Styles/sec/intro/intro}
\section{Method}

\input{Styles/sec/method/model}

\section{Data Cycling System}
\label{data_chapter}

\input{Styles/sec/data/main_data}

\section{Experiments}
\input{Styles/sec/exp/main_exp}

\section{Conclusion and Future Work}
\input{Styles/sec/conclusion/con}

\bibliographystyle{plainnat}
\bibliography{ref}


\end{document}

%% file: commands.tex
\newcommand{\SamJ}[1]{\textcolor{red}{#1}}
\newcommand{\negHL}[1]{\textcolor{red!60}{#1}}
\newcommand{\posHL}[1]{\textcolor{blue!50}{#1}}

\newcommand{\tablestyle}[2]{\setlength{\tabcolsep}{#1}\renewcommand{\arraystretch}{#2}\centering\footnotesize}
\renewcommand{\paragraph}[1]{\vspace{1.25mm}\noindent\textbf{#1}}
\newcommand\blfootnote[1]{\begingroup\renewcommand\thefootnote{}\footnote{#1}\addtocounter{footnote}{-1}\endgroup}

\newcolumntype{x}[1]{>{\centering\arraybackslash}p{#1pt}}
\newcolumntype{y}[1]{>{\raggedright\arraybackslash}p{#1pt}}
\newcolumntype{z}[1]{>{\raggedleft\arraybackslash}p{#1pt}}

\newcommand{\app}{\raise.17ex\hbox{$\scriptstyle\sim$}}
\newcommand{\mypm}[1]{\color{gray}{\tiny{$\pm$#1}}}
\newcommand{\x}{{\times}}
\definecolor{deemph}{gray}{0.6}
\newcommand{\gc}[1]{\textcolor{deemph}{#1}}
\definecolor{baselinecolor}{gray}{.9}
\newcommand{\baseline}[1]{\cellcolor{baselinecolor}{#1}}
\newcommand{\authorskip}{\hspace{2.5mm}}

\definecolor{dt}{HTML}{ADCAD8}
\definecolor{dt2}{HTML}{cddfe7}
\newcommand{\ots}[1]{\textcolor{dt}{#1}}
\newcommand{\otsmodel}[1]{\cellcolor{dt2}{#1}}

\definecolor{defaultcolor}{HTML}{E8E2F7}
\newcommand{\default}[1]{\cellcolor{defaultcolor}{#1}}

\let\cite\citep

\newcommand{\dslope}{\text{\ding{216}}}%
\newcommand{\fslope}{\text{\ding{217}}}%
\newcommand{\uslope}{\text{\ding{218}}}%

\newcommand{\scolorbox}[2]{{\setlength{\fboxsep}{2pt}\colorbox{#1}{#2}}}

\renewcommand{\paragraph}[1]{\vspace{1.25mm}\noindent\textbf{#1}}
\newlength\savewidth\newcommand\shline{\noalign{\global\savewidth\arrayrulewidth
  \global\arrayrulewidth 1pt}\hline\noalign{\global\arrayrulewidth\savewidth}}
\newcommand\hshline{\noalign{\global\savewidth\arrayrulewidth
  \global\arrayrulewidth 0.5pt}\hline\noalign{\global\arrayrulewidth\savewidth}}
\definecolor{degray}{gray}{.6}
\newcommand{\deemph}[1]{\textcolor{degray}{#1}}

%% file: Styles/sec/intro/intro.tex
In the digital world, graphical user interfaces (GUIs) serve as the primary gateway for human-computer interactions, permeating everyday activities such as web browsing, mobile app usage, and software navigation. With users spending an increasing portion of their time on digital devices, autonomous GUI agents—intelligent systems capable of perceiving, reasoning, and acting within GUI environments—hold immense potential to automate repetitive tasks, enhance accessibility, and streamline workflows. For instance, these agents can facilitate complex operations like e-commerce searches, form submissions, or multi-step interactions across platforms, thereby boosting efficiency in domains from personal productivity to enterprise automation. Recent advancements in large language models (LLMs) and visual language models (VLMs) have accelerated progress in this area, enabling agents to interpret screenshots and execute actions in a human-like manner, as demonstrated in applications such as web navigation~\cite{hong2024cogagent} and device control~\cite{wang2024survey,nguyen2024gui}.

Despite these advances, existing approaches to GUI agents exhibit both strengths and limitations. Many prior GUI agents adopt a modular hybrid approach, facilitating quick domain-specific task development through text extraction~\cite{zeng2025bridging}, understanding and reasoning modules~\cite{hurst2024gpt-4o}, and memory storage modules~\cite{xie2024osworld}. However, this dependence on expert input, and specialized VLMs renders them vulnerable to failures from minor shifts in tasks, VLMs, or environments~\cite{xia2024agentless}, yielding poorer scalability and adaptability than modern end-to-end frameworks. On the positive side, methods leveraging VLMs, such as those built on models like Qwen-VL~\cite{qwenvl,Qwen2VL} or CogVLM~\cite{wang2024cogvlm}, offer versatility by directly processing visual inputs (e.g., screenshots), eliminating the need for structured text like HTML or APIs. This visual-centric paradigm, exemplified in CogAgent~\cite{hong2024cogagent} and GUICourse~\cite{chen2024guicourse}, enables robust handling of diverse GUI elements, including icons, buttons, images, and spatial layouts, and has achieved state-of-the-art performance on benchmarks like Mind2Web~\cite{deng2023mind2web} and AITW~\cite{rawles2023androidinthewild}. Recent advancements in VLM-based GUI agents have extended to specialized domains, such as mobile environments. For instance, MagicGUI~\cite{tang2025magicgui} proposes a foundational agent for mobile GUIs, leveraging a scalable data pipeline with continued pre-training (CPT) and reinforcement fine-tuning to enhance perception and grounding in high-density smartphone interfaces. However, while MagicGUI focuses on mobile-specific challenges such as swipe gestures and app-centric navigation, our Mano framework targets web and desktop GUIs, incorporating unique components like the Mano-parking module for autonomous data extraction and Mano-verify for error recovery. This differentiation allows \textbf{Mano} to address broader cross-platform variability, including dynamic web structures and operating system diversity. Reinforcement learning (RL)-based fine-tuning~\cite{Leo2024AgentNet, feng2025towards,zhang2025rearank}, as explored in DigiRL~\cite{bai2024digirl}, GUI-RL~\cite{luo2025gui}, and WebRL~\cite{qi2024webrl}, further enhances decision-making in multi-step trajectories, narrowing gaps in purely supervised fine-tuning(SFT). However, these methods often suffer from drawbacks: \textbf{(1)} reliance on low-resolution processing, which leads to inaccuracies in recognizing fine-grained elements like small text or icons; \textbf{(2)} domain-limited or simplistic datasets that fails to capture real-world variability and stochasticity across various operating systems and websites; \textbf{(3)} LLMs/VLMs tuned via SFT alone may excel in single-step predictions but lack the holistic reasoning needed for complete interaction sequences, resulting in suboptimal performance in dynamic environments.

These limitations underscore persistent challenges in GUI agent development. \raisebox{-0.5pt}{\ding[1.1]{182\relax}} Data mismatch: pre-trained VLMs~\cite{qwenvl, Qwen2VL, CogVLM2} are often optimized for natural images rather than GUI-specific content, leading to poor optical character recognition (OCR), grounding, and widget understanding ability in complex interfaces. \raisebox{-0.5pt}{\ding[1.1]{183\relax}} Inefficient decision-making in long-horizon tasks: as SFT objectives focus on immediate predictions without rewarding end-to-end success, while RL-based exploration—especially online—can be computationally expensive and susceptible to policy drift. \raisebox{-0.5pt}{\ding[1.1]{184\relax}} Sim-to-Real gap: The inefficiency inherent in human-annotated trajectory collection, coupled with the sparsity of training data, results in limited model adaptability to real-world environmental variations, including dynamic events, operating system changes, and user interface element updates.

To address these challenges, we introduce \textbf{Mano}, a web GUI agent built upon a multi-modal foundation model pre-trained on extensive web data. Specifically, our base model is UITARS-1.5-7B~\cite{qin2025ui}, derived from Qwen2.5-VL-7B~\cite{bai2025qwen25vltechnicalreport} through RL fine-tuning on GUI-related data. A key innovation is our custom-designed simulated environment, which efficiently generates high-quality interaction data from diverse operating systems, enabling robust data collection for all training stages while mitigating real-world deployment costs and variability. 

\begin{figure}[t]
    \centering  
    \includegraphics[width=1.0\linewidth]{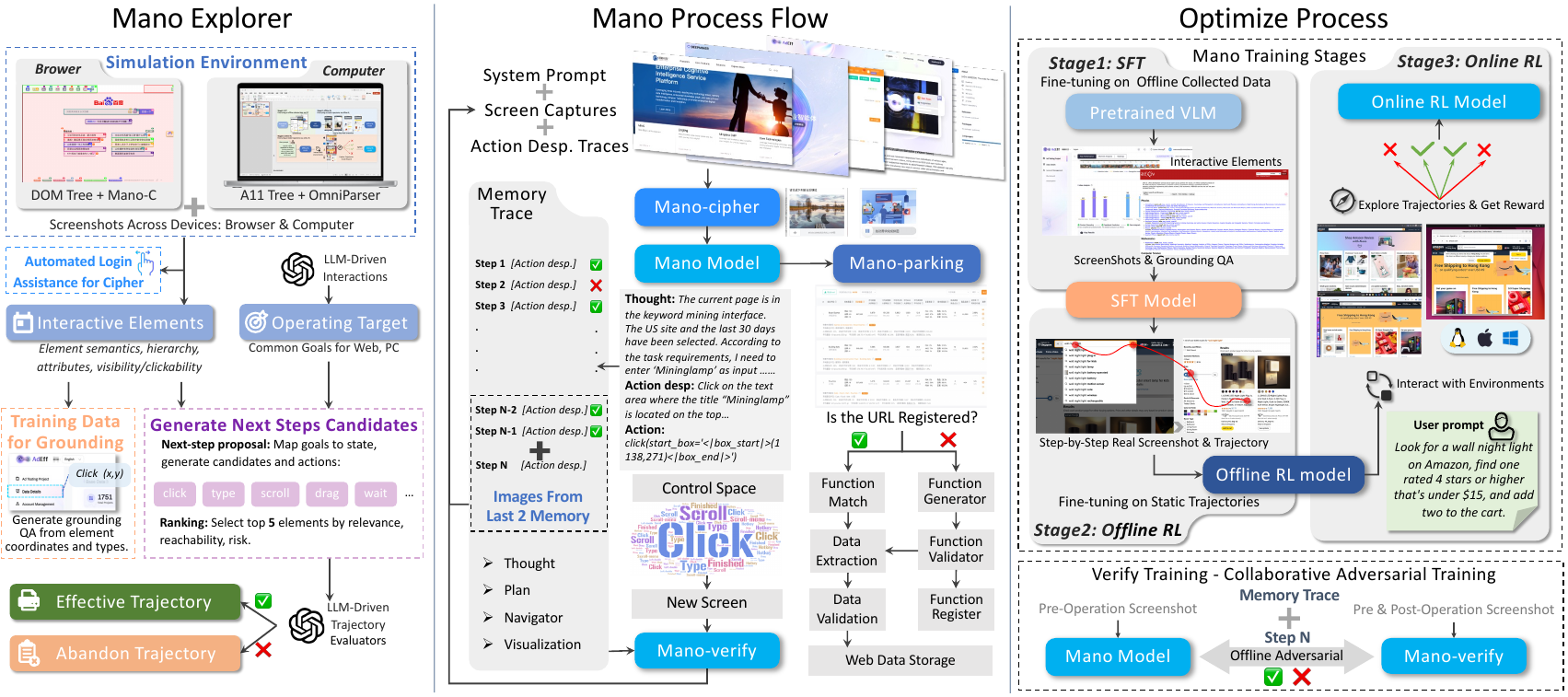}
    \caption{Overview of the Mano framework. The left part illustrates the \textit{Exploration Module}, which operates in simulated browsers and desktop environments to collect interaction elements and candidate goals, generating diverse trajectories and login assistance data for training. The center shows the \textit{Inference Process Pipeline}, where the model follows a structured “think–act–verify” loop: interpreting GUI states, producing action descriptions (e.g., clicks or type), executing them, and validating outcomes through a verifier. The right part depicts the \textit{Optimize Process}, a progressive pipeline of SFT, offline RL, and online RL, which systematically strengthens reasoning, adaptability, and end-to-end decision-making in dynamic GUI environments.}
    \label{fig:agent_framework}
    \vspace{-8mm}
\end{figure}

As illustrated in Fig.~\ref{fig:agent_framework}, our framework consists of three tightly coupled components. On the left, an \textbf{exploration module} operates in simulated browsers and desktop environments to collect interaction primitives and candidate goals, forming diverse trajectories for downstream training. This module also features automated login assistance that not only manages cipher tasks but also automatically collects related data, facilitating access to secure systems without compromising security protocols. In the center, the \textbf{inference pipeline} of \textbf{Mano} drives task execution through a structured loop of thinking–acting–verifying: the model interprets screenshots and prompts, generates action descriptions (e.g., clicks or inputs), executes them as concrete commands, and leverages a verifier to ensure consistency and recover from errors. The inclusion of the Mano-cipher enhances the system's capability to handle intricate tasks requiring secured data entry. Mano-parking is utilized for crawling both structured and unstructured data from web pages, providing valuable input for downstream processes. On the right, the \textbf{training process} integrates these signals into a progressive pipeline of SFT, offline RL and online RL, aligning static knowledge with robust multi-step decision-making in dynamic GUI environments.  

In the first stage, we perform full-parameter SFT on the model using carefully processed interaction data sourced from real data and simulated environment across multiple websites and operating systems. This stage enables the model with focused, accurate contextual understanding (detailed data processing methods are described in Sec.~\ref{data_chapter}), producing an initial model denoted as \textbf{Mano-SFT}. However, this SFT-tuned model still lack the end-to-end decision-making capability required for complete GUI interaction trajectories, as the SFT objective only requires predicting the current step’s reasoning, summary and action based on preceding context.

To bridge this gap, the second stage employs offline RL fine-tuning with  group relative policy optimization (GRPO)~\cite{shao2024deepseekmathpushinglimitsmathematical}, leveraging data from the simulated environment and designing rewards specifically tailored to encourage successful completion of full interaction sequences offline, enhancing the model's holistic GUI reasoning and GUI decision-making ability. This yields \textbf{Mano-Off}, a more capable intermediate model.

Finally, to further adapt the model to diverse operating environments and GUI interactions, we deploy online RL in the third stage based on our simulated environment, again leveraging GRPO but with distinct rewards focused on real-time adaptability and exploration in dynamic settings. During this phase, the agent collects new interaction data through online trials, which is then cycled back as offline data for further refinements, enabling continuous improvement via iterative loops. This culminates in the final \textbf{Mano}, which demonstrates superior robustness across varied web GUI scenarios.

Through this progressive pipeline, Mano directly addresses the challenges: it resolves data mismatches by integrating domain-specific, high-fidelity simulated interactions; strengthens multi-step reasoning via targeted RL rewards; and ensures adaptability through efficient online exploration in a controlled yet diverse environment. Our contributions encompass this comprehensive training framework, empirical validations on GUI benchmarks, and novel insights into RL's application—particularly GRPO with stage-specific rewards—in overcoming VLM limitations for practical agent deployment.

To summarize, \textbf{Mano} presents the following contributions:
\begin{enumerate}

    \item A novel simulation environment for iterative data generation: we design an simulated environment that efficiently produces high-quality interaction data from diverse operating systems and various websites, supporting all training stages through a cyclical process—where online exploration collects new data that feeds back into offline updates—thus addressing data mismatches and scalability issues while reducing real-world deployment costs.

    \item Novel insights on RL integration: through empirical evaluations on GUI benchmarks, we provide new insights into RL's role, including a tailored application of the GRPO algorithm with distinct reward functions for the offline and online RL stages—balancing policy stability with goal-oriented optimization—in bridging limitations of VLMs for practical, adaptable GUI agent deployment.
    
    \item State-of-the-art performance: Mano achieves state-of-the-art (SOTA) results on multiple GUI benchmarks, outperforming prior agents in metrics like success rate and efficiency, validating the effectiveness of our integrated framework in real-world GUI navigation and interaction tasks. 
\end{enumerate}

%% file: Styles/sec/method/model.tex
\begin{figure}[h]
    \centering   \includegraphics[width=1.0\linewidth]{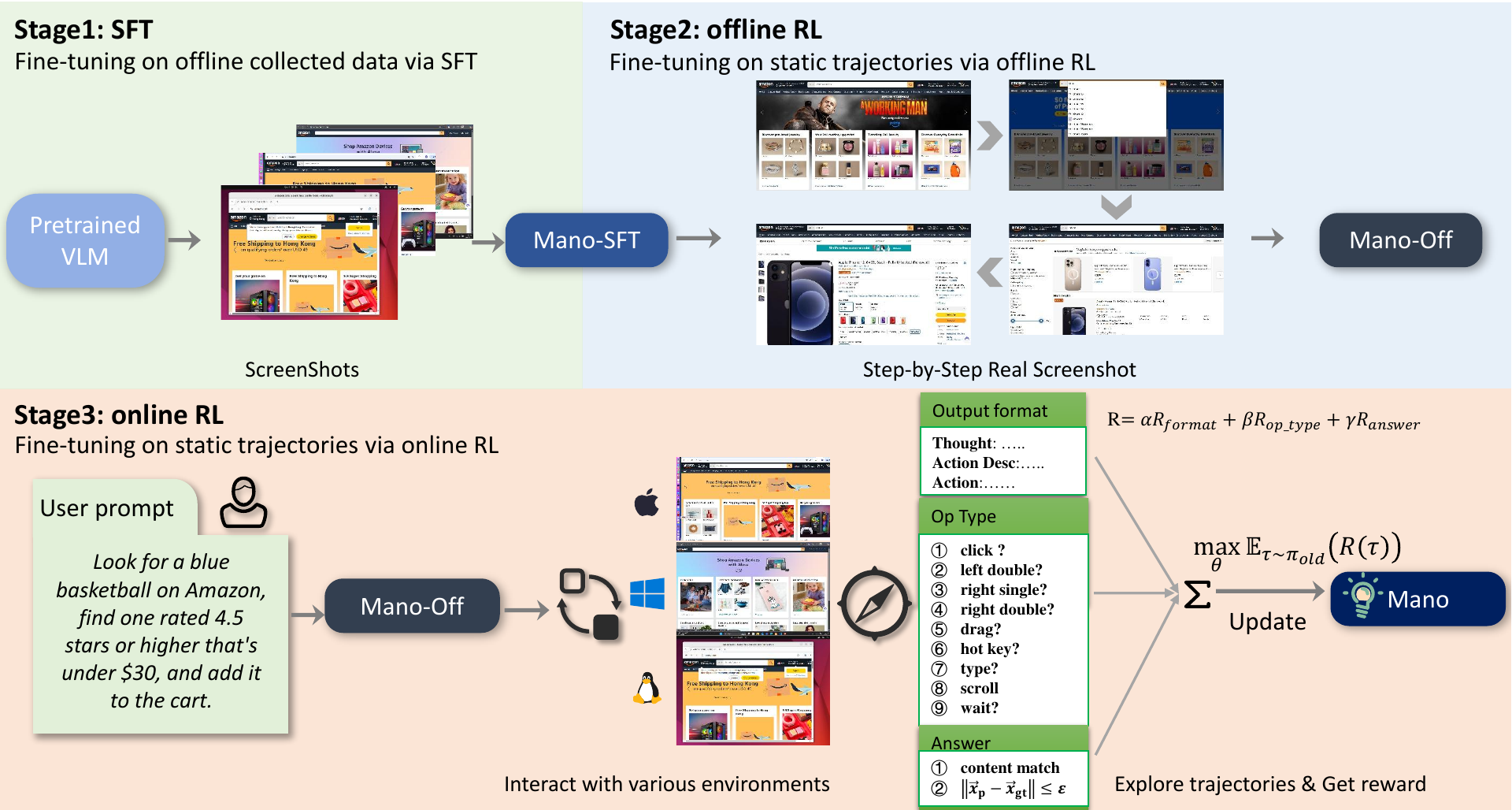}
    \caption{Overall fine-tuning framework of \textbf{Mano} for GUI-oriented tasks. The pipeline consists of three progressive stages: (i) \textbf{SFT} on offline demonstrations; (ii) \textbf{Offline RL} leveraging static trajectories with reward decomposition; and (iii) \textbf{Online RL} with active environment interaction. The system incorporates step-level reasoning, explicit action description, and operation type selection (e.g., click, drag, type, scroll), while final performance is evaluated through structured outputs and multi-dimensional rewards combining format accuracy, operation correctness, and task completion.
    }
    \label{fig:train_framework}
    \vspace{-4mm}
\end{figure}
\subsection{Training of Mano}
Our Mano is built upon a multimodal foundation model pre-trained on extensive web data. Specifically, our base model is UITARS-1.5-7B~\cite{qin2025ui}, which is derived from Qwen2.5-VL-7B via RL fine-tuning using GUI-related data. As illustrated in Fig.~\ref{fig:train_framework}, the overall framework of \textbf{Mano} proceeds through three consecutive learning stages: SFT, offline RL and online RL fine-tuning. Throughout the entire training process, we employ full parameter fine-tuning.

\subsubsection{First-stage: Supervised Fine-tuning}
The primary objective of this initial stage is to bridge the domain gap between the general-purpose pre-training of the base VLM and the specialized requirements of web GUI interaction. This phase is designed to instill a robust perceptual and semantic foundation in the model, specifically tailored to the unique visual grammar of computer interfaces.

To achieve this, our base VLM is fine-tuned on a curated dataset of decision-centric GUI interaction trajectories, sourced from multiple websites across different operating systems. Details of data processing will be elaborated in Sec.~\ref{data_chapter}. A critical aspect of our data processing is the preservation of native, dynamic image resolution. In line with the methodology of Qwen2.5-VL, we avoid aggressive downsampling. 

This is predicated on the observation that GUI screens are highly sensitive to fine-grained details; elements such as small font text, subtle icons, buttons, text, and diverse layout structures (which can be highly dense), dialogs, or densely packed widgets are prevalent in real-world applications and are often imperceptible at lower resolutions. Maintaining high fidelity in the visual input is therefore paramount for accurate perception and grounding.

While parameter efficient fine-tuning (PEFT) methods, such as LoRA~\cite{hu2022lora,zhou-etal-2025-gsq} or adapter tuning~\cite{han2024parameter,Zhang2025SensitivityLoRA,Zhang2025TimeLLaMA}, are effective for adapting models to new tasks or domains, we posit that they are insufficient for rectifying the fundamental domain mismatch between natural images and GUIs. Adapting the model to the unique characteristics of GUIs—including its distinct OCR patterns, widget affordances, and spatial semantics—requires substantial updates to the model's core components, particularly the vision encoder and the cross-modal attention layers.

Therefore, we perform \textbf{full-parameter SFT}, unfreezing the vision-language adapter and the language model while keeping the visual backbone frozen, to allow for a comprehensive adaptation of its internal representations to the target domain. Formally, given a dataset of expert trajectories $\mathcal{D}_{\text{SFT}}$, where each trajectory $\tau = \{(s_1, y_1), (s_2, y_2), \dots\}$ consists of a sequence of states and the corresponding expert utterances (containing reasoning, summary and action), our objective is to maximize the log-likelihood of these expert utterances. The SFT loss function is defined as the standard auto-regressive cross-entropy loss over the tokens of the target utterances:

\begin{equation} 
\label{eq:sft_loss}
\mathcal{L}_{\text{SFT}} = - \mathbb{E}_{\tau \sim \mathcal{D}_{\text{SFT}}, (s_i,y_i) \sim \tau} \log P(y_{i,j} | s_i, y_{i,<j}; \theta)
\end{equation}

where $\theta$ represents the full set of model parameters, $s_i$ is the multi-modal state input at step $i$, and $y_{i,j}$ is the $j$-th token of the target expert utterance $y_i$. By optimizing this objective across all model parameters, we encourage a deep, foundational alignment with the GUI domain, resulting in the \textbf{Mano-SFT} model, which serves as a highly capable starting point for the subsequent reinforcement learning stages.

\subsubsection{Second-stage: Multi-step Reasoning via Offline Reinforcement Learning} 
While the \textbf{Mano-SFT} model acquires a strong perceptual foundation, its training objective—maximizing the likelihood of the next expert action—predicting the current step’s reasoning, summary, and action based on preceding context. This single-step focus does not guarantee optimal performance over long, multi-step interaction trajectories. As shown in Tab.~\ref{tbl:stage_ablation}, the Mano-SFT model achieves a score of only 32.7 on the OSWorld-Verified~\cite{xie2024osworld} benchmark.

To bridge the gap between single-step accuracy and holistic, multi-step reasoning, we introduce an offline reinforcement learning stage. This phase allows the agent to learn from the outcomes of entire trajectories without the high cost and potential instability of live, online exploration. We fine-tune the \textbf{Mano-SFT} model using GRPO , rewarding the model for correctly completing full interaction sequences. This step enhances overall task completion ability. To further adapt the model to diverse operating environments and dynamic GUI interactions, we deploy it in a simulated environment for online RL. During this stage, we freeze the adapter while keeping the LLM part trainable.

\paragraph{Reward Design for Offline RL.}
In the offline setting, it is crucial to encourage the agent to learn effective strategies while preventing its policy from deviating drastically from the reliable expert data. Therefore, we design a dense, process-oriented reward functionx` that provides granular feedback at each step. 

Specifically, \textbf{Mano} can be formulated as a finite-horizon Markov Decision Process (MDP)~\cite{sutton2018reinforcement}, defined by the tuple \( \mathcal{M} = \{ S, \mathcal{S}_0, \mathcal{A}, \mathcal{R}, \mathcal{T}, H \} \), where:

\begin{itemize}
    \item \(\mathcal{S}\) denotes the state space, incorporating user inputs and screenshots;
    \item \(S_0\) defines the initial state, determined by the environment,varied across different operating systems and websites;
  \item \(\mathcal{A}\) represents the finite action space (i.e., Mano's actions);
  
  \item \(\mathcal{T}: \mathcal{S} \times \mathcal{A} \times \mathcal{S} \to [0,1]\) specifies the transition probability to the next state given current state and action;
  \item \(\mathcal{R}: \mathcal{S} \times \mathcal{A} \to \mathbb{R}\) is a rule-based reward function for task completion;
  \item \(H\) indicates the finite steps of the episode per task.
\end{itemize}
The objective during RL stage is to maximize the cumulative reward.

\begin{equation}
\max_{\theta} \mathbb{E}[\sum_{i=0}^{H-1}R_i(s_i,a_i)]
\end{equation}

where $R_i(s_i,a_i) \in \mathbf{R}$ denotes the reward at step $i$, formulated as a weighted sum of three components: a format reward($R_{\text{format}}$), a type reward($R_{\text{op\_type}}$), and a final answer reward($R_\text{answer}$). The reward $R$ is a weighted sum of three components:
\begin{equation} \label{eq:reward_offline}
    R = \alpha R_{\text{format}} + \beta R_{\text{op\_type}} + \gamma R_{\text{answer}}
\end{equation}
where $R_{\text{format}}$ provides a positive reward if the agent's generated utterance conforms to the required format, $R_{\text{op\_type}}$ rewards the selection of a plausible action type given the context, and $R_{\text{answer}}$ denotes the reward based on either spatial criteria (i.e., whether the result falls within the GT bounding box or maintains a distance to the target point below a given threshold) or textual matching accuracy. The weights are set to prioritize correctness and progress while penalizing invalid outputs ($\gamma > \beta > \alpha$). We omit $s_i,a_i$ for simplicity and $\alpha, \beta,\gamma \in (0,1), \alpha+\beta+\gamma = 1$. This dense reward structure provides a stable learning signal that refines the agent's reasoning process step-by-step.

The optimization objective follows the GRPO formulation, which normalizes rewards within a group of trajectories sampled for the same task to generate advantages and clips the probability ratio to constrain policy updates:
\begin{equation} \label{eq:grpo_loss}
\mathcal{L}^{\text{GRPO}} = -\mathbb{E}_{\tau \sim \mathcal{D}_{\text{offline}}} \left[ \sum_{t=0}^{H-1} \min \left( \frac{\pi_\theta(a_t|s_t)}{\pi_{\text{ref}}(a_t|s_t)} \hat{A}_t, \text{clip}\left(\frac{\pi_\theta(a_t|s_t)}{\pi_{\text{ref}}(a_t|s_t)}, 1-\epsilon, 1+\epsilon\right) \hat{A}_t \right) \right]
\end{equation}
where $\hat{A}_t$ is the normalized advantage estimated from the group-wise rewards.

\begin{figure}[h]
    \centering   
    \includegraphics[width=1.0\linewidth]{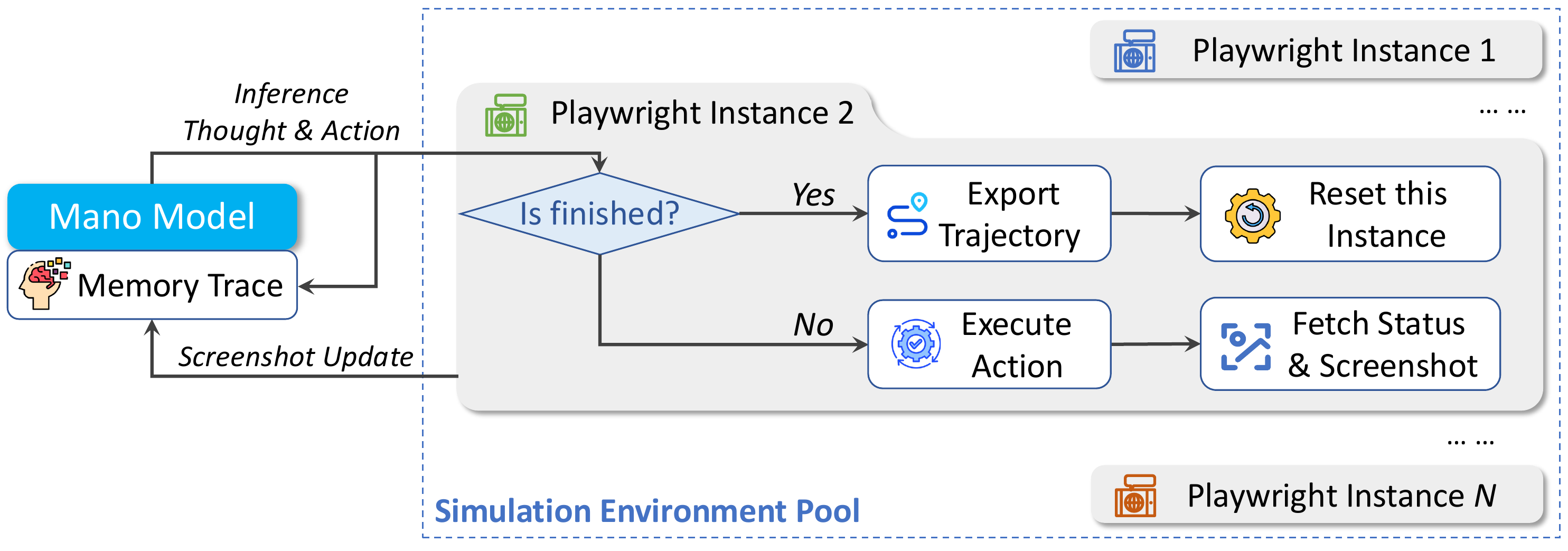}
    \caption{Overall framework of online RL in Mano. 
    The Mano model interacts with multiple parallel \texttt{Playwright} instances, each representing a GUI environment. 
    For every step, the model fetches the status and screenshot, performs inference to generate thought and action, and then executes the action within the corresponding environment. The loop continues until the task is completed, while memory traces are recorded and trajectories are exported for further training and analysis.}
    \label{fig:online_RL}
\end{figure}

\subsubsection{Third-stage: Online Reinforcement Learning} 
As shown in Fig.~\ref{fig:online_RL}, during the online RL phase, we establish our own simulation environment pool for model interactions with real environments. For browser-use agent (BUA) environments, we launch a group of browsers and communicate with pre-opened browsers through Chrome DevTool protocol (CDP) of Playwright\cite{Feldman2025microsoft}, each instance of which is assigned to one browser. By managing this connection pool, we enable simultaneous operations across multiple browsers. For computer-use agent (CUA) environments, we utilize Docker containers with Ubuntu images, implementing multi-environment concurrent model interactions through Docker instance pool management.

During the online RL phase, each batch contains only user prompts. Upon training initiation, each training task in the batch launches a virtual environment to retrieve current environmental information, such as screenshots. This environmental information is fed to the model for prediction, and the predicted actions are parsed into actual operations applied to the environment. Through this interaction, we obtain operation trajectories and corresponding completion statuses for each training task in real environments. This online sampling approach captures more environmental variations, compensating for the sparsity of offline trajectory distributions.

We do not directly employ online interaction with model updates during training due to the high temporal cost of such interactions. The approach of online sampling with offline filtering enables more effective cleaning and filtering of noisy trajectories, while allowing the implementation of various strategies to adjust the difficulty distribution of trajectory samples, preventing ineffective learning from excessive failure trajectories.

Following this phase, as shown in Tab.~\ref {tbl:stage_ablation}, the model's average score on the OSWorld-Verified dataset~\cite{xie2024osworld}  improves by 7.9, reaching 41.6. The three aforementioned stages can be iterated cyclically until performance improvements reach a saturation point on our validation set. Through this training process, we obtain the final Mano model.

\subsection{Mano-parking}

\begin{figure}[h]
    \centering   
    \includegraphics[width=1.0\linewidth]{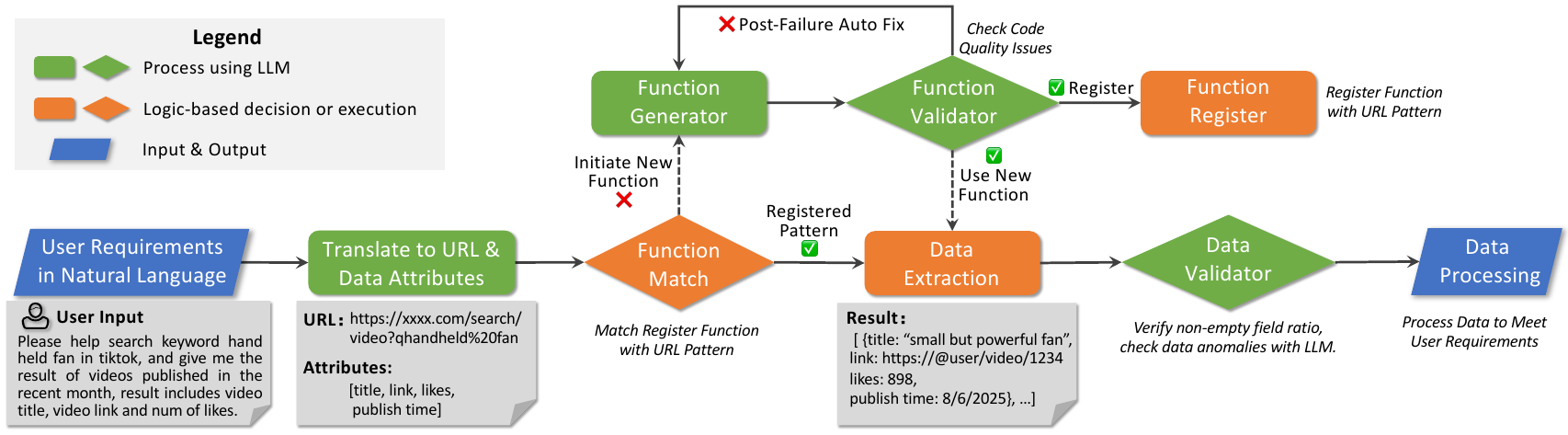}
    \caption{
    The operational workflow of \textbf{Mano-parking}, which illustrates its autonomous data extraction pipeline. The process begins with request reception and function registry lookup, followed by either direct execution of pre-validated functions or initiation of a multi-phase extraction synthesis. In the latter case, simplified HTML structures are obtained through browser automation and cleaning algorithms, combined with user-defined attribute specifications to generate customized extraction functions. These functions undergo a three-tier validation—field completeness, semantic consistency, and structural integrity—before being executed and stored for reuse. Furthermore, Mano-parking incorporates continuous monitoring and a self-healing mechanism, enabling adaptive regeneration of extraction logic when website structures evolve. This design ensures robustness, efficiency, and minimal human intervention across diverse web environments.
    }
    \label{fig:mano-parking}
\end{figure}

Within the Mano ecosystem, Mano-parking represents a breakthrough in autonomous data extraction technology. This specialized component transforms unstructured web content into organized, actionable datasets without requiring users to possess programming expertise. Unlike conventional web scrapers that require constant maintenance, Mano-parking functions as an intelligent agent that understands, adapts to, and extracts data from diverse website architectures with minimal human oversight. The overall workflow of Mano-parking is illustrated in Fig.~\ref{fig:mano-parking}, highlighting its automated pipeline from request handling to adaptive self-repair.

The operational workflow of Mano-parking follows a deterministic and efficiency-oriented protocol. 
Upon receiving an extraction request, the system queries its function registry using the target URL as the primary identifier. In cases where a corresponding extraction function exists within the registry, the system executes the pre-validated function with minimal computational resources. Conversely, when the system encounters an unregistered URL pattern, it initiates a multiphase extraction synthesis process: the system acquires semantically preserved, structurally simplified HTML content through browser automation and intelligent cleaning algorithms, which it then integrates with user-specified natural language attribute requirements to generate a tailored extraction function. The newly generated function subsequently undergoes systematic validation before execution and registration in the function repository for subsequent utilization.

To guarantee data quality and reliability, Mano-parking incorporates a comprehensive three-tiered validation framework.
The first tier verifies extraction completeness by monitoring required and optional field coverage against domain-specific thresholds. The second tier employs language models to detect semantic anomalies, identifying logical inconsistencies and contextual irregularities that traditional validation methods might miss. The third tier analyzes the structural integrity of the code, examining generated extraction functions for proper syntax, resource utilization, exception handling, and architectural soundness to prevent potential run-time failures.

Autonomous self-evolution distinguishes Mano-parking from conventional extraction systems. The platform continuously monitors extraction function performance through scheduled health checks, detecting when website structural changes -- inevitable in today's rapidly evolving digital landscape -- cause function degradation. Upon validation failure, the auto-repair module analyzes the failure context, examines the current website structure, and autonomously regenerates extraction logic, adapting to structural modifications while maintaining extraction intent. This self-healing capability dramatically reduces maintenance overhead and ensures consistent data availability despite the evolution of the source website.

Besides this integration of validation frameworks and self-correction mechanisms, the system's core technical innovations include: (1) browser automation for navigating websites, circumventing risk control mechanisms, executing preprocessing actions, and loading dynamic content;(2) intelligent HTML cleaning algorithms achieving >90 percent compression with particularly high efficiency for JavaScript-heavy websites; (3) adaptive prompt engineering for context-optimized code generation; (4) component-based URL pattern recognition for precise function matching.

\subsection{Mano-verify Model}

Within the Mano system, Mano-verify serves as a crucial verification module designed to ensure the correctness of every step of planning and execution. While the Mano model is responsible for generating actions across multi-modal contexts, Mano-verify functions as a safeguard, providing an independent assessment of whether a given operation has been carried out accurately. Its primary role is to judge the fidelity of action execution, detect possible discrepancies, and immediately intervene in the reasoning loop when errors are detected. This verification mechanism enhances the robustness of the entire system, preventing error propagation and enabling self-correction.

The inputs to Mano-verify are carefully structured to capture the multimodal nature of interaction. They include: (i) the pre-operation screenshot, representing the state of the interface before execution; (ii) the post-operation screenshot, reflecting the environment after the action; and (iii) textual context, consisting of the system prompt, the action description produced by the Mano model, and a shared history of previous operations. By jointly reasoning over these visual and textual signals, the verification model determines whether the current step has been executed correctly.

Training Mano-verify involves a distinct adversarial element, reflecting the dynamic interplay between action description and evaluation. The foundation of the training corpus is built from two complementary sources: curated data collection and simulation environments. These datasets predominantly yield positive samples, representing correct operations with clear instructional value. However, reliance on positive examples alone would bias the verifier toward overconfidence. To address this, we systematically harvest trajectories from failed tasks during real Mano runs. These failed paths, rich in incorrect operations, are then corrected through human-in-the-loop intervention. The resulting dataset not only supplies valuable negative samples for supervision but also provides revised action descriptions that feed back into the training of the Mano model itself. In this way, Mano-Verify supports a co-evolution of planning and validation, strengthening both components of the system.

Formally, we denote the input state as multimodal tuple
\begin{equation}
x_t = \{ I^{pre}_t, I^{post}_t, p_t, a_{desp.}, h_t \},
\end{equation}

where $I^{pre}_t$ and $I^{post}_t$ represent pre- and post-operation images, $p_t$ is the system prompt, $a_{desp.}$ is the last action description, and $h_t$ denotes the historical trace. The verification model outputs

\begin{equation}
y_t = f_\theta(x_t),
\end{equation}

where $y_t \in \{\text{correct}, \text{incorrect}\}$, along with diagnostic labels pointing to errors in either description or execution. 

During inference, Mano-verify operates as a stepwise checkpoint. After each Mano model, it evaluates correctness and records the judgment within the system’s memory trace. This feedback loop is deliberately lightweight and expressive: verification outcomes are stored not only as structured annotations but also marked with intuitive emoji symbols, signaling success or failure. Such symbolic cues, embedded in the history available to the Mano model, function as compact yet powerful indicators of operational reliability. The process can be abstractly written as:

\begin{equation}
h_{t+1} = h_t \oplus \text{verify}(x_t),
\end{equation}

where $\oplus$ denotes the augmentation of history with verification feedback.

This ensures that the agent’s decisions remain accurate, interpretable, and aligned with intended goals, ultimately enhancing the practicality and trustworthiness of Mano in real-world applications.

\begin{figure}[h]
    \centering   \includegraphics[width=1.0\linewidth]{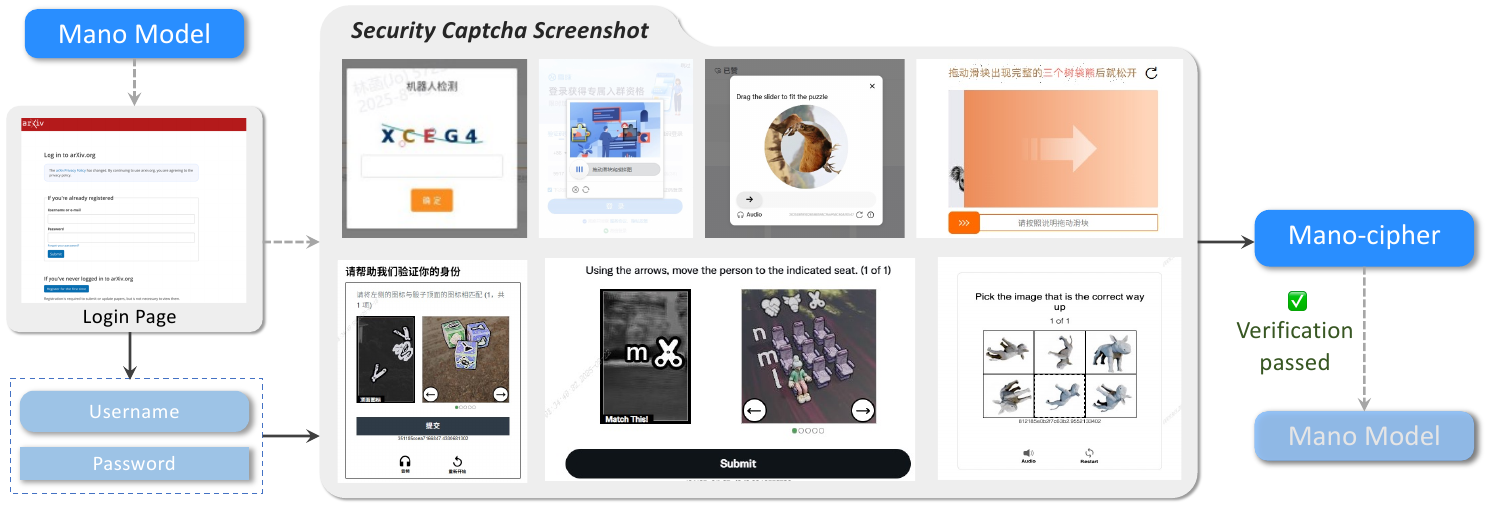}
    \caption{Mano-cipher is a specialized authentication GUI model. This GUI model facilitates automated login operations across diverse systems by handling various captcha types—including alphanumeric, image-based sliding, rotation, content recognition, and logical reasoning challenges. Upon successful verification, system control is returned to the Mano for subsequent tasks.
    }
    \label{fig:mano-cipher}
    \vspace{-4mm}
\end{figure}

\subsection{Mano-cipher}
Mano-cipher represents an advanced authentication graphical user interface (GUI) model that seamlessly integrates credential verification with the Mano online RL mechanism. The primary objective of Mano-cipher is to facilitate automated login processes for diverse authentication systems, encompassing the input of usernames and passwords, as well as the completion of various verification code challenges. As depicted in Fig.~\ref{fig:mano-cipher}, when encountering a system or web interface that requires user authentication, Mano asks the user for their credentials, subsequently automating the completion of the form and navigating through multiple verification code procedures. Mano-cipher exhibits robust capabilities in managing an array of verification code modalities, including but not limited to alphanumeric sequences, image-based sliding puzzles, image rotation tasks, content recognition challenges, and logical reasoning tests. Upon successful authentication, Mano-cipher efficiently transfers system control back to Mano, enabling the continuation of subsequent operational tasks.

In the Mano-cipher training pipeline, we leverage a dataset of 2,000 landscape images as backgrounds for synthesizing CAPTCHAs, generating paired training images and ground-truth operational sequences for diverse CAPTCHA instances. The training proceeds in two stages: (1) We first apply SFT to establish an initial model that produces properly formatted actions with acceptable coordinate localization accuracy; (2) We then implement online RL by deploying multiple headless browser instances concurrently to generate varied CAPTCHAs as online training data, utilizing rule-based reward signals for GRPO. Through iterative real-time environment interactions within this reinforcement learning framework, we obtain the final optimized model.

%% file: Styles/sec/data/main_data.tex
\input{Styles/sec/data/data_template}
\input{Styles/sec/data/space_def}

\subsection{Template and Action definition}
To train Mano, we design a concise template to guide the model in following specific instructions. As depicted in Tab.~\ref{tbl:template}, this template first requires the model to reason about the user’s input, then provide a succinct summary(For simplicity, we use summary to refer to 'Action Desp'. Henceforth, summary and 'Action Desp' are used interchangeably in the text.) of the reasoning process, and finally generate an answer that must fall within a predefined action space given in Tab.~\ref{tbl:action_def}.

\input{Styles/sec/data/fig/fig_data_engine}
\subsection{Unified Trajectory Collection Pipeline for Desktop and Web Environments}
To efficiently collect interaction trajectories across both desktop and web environments, we develop a unified automated data collection pipeline with domain-specific adaptations. As shown in Fig.~\ref{fig:data_framework}, the detailed procedure is exemplified through the workflow of collecting operational trajectories in web environments.
\begin{enumerate}
\item \textbf{Infrastructure and Objective Generation}
We establish a scalable cluster of virtual environments capable of simulating diverse interaction scenarios. For each target application—whether a web URL or desktop software module—we employ Claude~\cite{Claude3} to automatically generate a prioritized list of functional objectives while filtering out rarely-used features. This curated objective list provides contextual guidance throughout the exploration phase.

\item \textbf{Multi-Modal Element Extraction}
For web environments, we develop a custom browser extension \textbf{Mano-C} that comprehensively extracts interactive elements, capturing both spatial coordinates and semantic attributes of each DOM element. This enables systematic collection of all interactive elements on web pages along with their bounding boxes and properties.
\textbf{Mano-C} is our proprietary Chrome extension designed for comprehensive extraction of interactive elements from web pages. The extension employs a systematic DOM tree traversal approach to identify: (1) HTML elements with explicit interactive semantics, (2) elements containing ARIA attributes, (3) elements with registered click event listeners, and (4) encapsulated interactive components. We implement a multi-tiered filtering pipeline that eliminates elements outside the viewport boundary and examines CSS properties including display, visibility, and opacity to exclude non-visible elements. Additionally, elements with negligible dimensions (e.g., 1×1 pixels)—commonly used for tracking purposes or as hidden elements—are filtered out. Our system incorporates specialized handling for modern web technologies, including recognition of Web Components and framework-specific custom components (React, Vue), detection of contentEditable regions, and identification of Canvas-based interactive regions through data attributes or designated class selectors. This comprehensive approach ensures accurate and exhaustive capture of all interactive elements present on web pages.
For desktop environments, we implement a hybrid approach combining Accessibility Tree (A11y Tree) parsing with OmniParse\cite{wan2024omniparser} for collaborative filtering, enabling robust extraction of interactive elements and their attributes across diverse desktop applications. This dual-mechanism approach ensures comprehensive coverage of UI elements that may be missed by single-method extraction.

\item \textbf{Element Annotation and Grounding}
We leverage large language models to annotate each extracted element with semantic labels, functional descriptions, and interaction categories, generating rich grounding data essential for training. This annotation process provides structured supervision signals for learning element-action associations.

\item \textbf{Intelligent Exploration Strategy}
We design a prompt-engineered module for strategic element selection during exploration, incorporating explicit constraints to prevent cyclic paths and redundant branch exploration. The exploration follows a depth-first search (DFS) strategy with a maximum depth of 10 levels, balancing thorough coverage with computational efficiency. At each explored state, the system captures screenshots and stores annotated interaction data for subsequent processing.

\item \textbf{Quality Assessment Pipeline}
Post-exploration, we implement a comprehensive trajectory scoring pipeline to identify high-quality interaction sequences. We formulate evaluation criteria as structured prompts, enabling Claude to assess trajectory quality across multiple dimensions including completeness, intent clarity, and task coherence. Only trajectories meeting stringent quality thresholds are retained, ensuring the final dataset comprises diverse, high-fidelity interaction demonstrations suitable for training robust agents.
This unified pipeline enables scalable collection of interaction trajectories across heterogeneous environments while maintaining consistency and minimizing manual annotation overhead.
\end{enumerate}

\input{Styles/sec/data/sft_data_orgs}
\subsection{Data organization}

Each instance of a GUI-agent task can be regarded as a complete trajectory consisting of N steps. Each step comprises a web screenshot, meta-information of web elements (such as functional descriptions and positions of buttons), click event coordinates, click response information, among other data. This type of data forms a multi-modal dataset integrating images, text, and actions. The data includes a system prompt $p_s$, a user prompt $p_u$, an image $o$ observed from the environment, the agent’s thinking process $t$, summary $s$, and the corresponding action $a$. Specifically, at SFT stage, 
each data sample is structured as Tab.~\ref{tbl:sft_data}, where $o_i,t_i,s_i,a_i$ represent the observation, thinking process, summary and corresponding action at the $i$-th step, respectively. For each data instance in the SFT stage, the input for the current step is constructed by retaining the historical observations from the previous two steps (if available), denoted as $o_{i-1},o_{i-2}$, along with the current observation $o_i$, as well as all historical summary records up to the current step. During the RL phase (including both offline and online RL), we directly provide the inputs $p_s,p_u,o_0$
 , expecting the model to leverage both its intrinsic knowledge and the prior experience gained during the SFT stage for decision-making in GUI environments to autonomously explore trajectories that maximize cumulative reward. This process is consistent with conventional RL frameworks.

\input{Styles/sec/data/fig/attention_map}
 
 During SFT stage, unlike the data organization scheme used in UI-TARS, we incorporate an additional summary following the thinking part at each step, which was absent in UI-TARS. Compared to the thinking process, we argue that a concise and clear summary exerts a more critical influence in guiding subsequent action generation. This approach also encourages the model to allocate greater attention to the succinct summary when generating actions. As shown in Tab.~\ref{tbl:sft_ablation}, this single modification in the data organization format alone leads to a performance improvement of 2.8 (increasing from 29.9 to 32.7). Concurrently, we visualize the attention maps of the Mano-SFT. As illustrated in Fig.~\ref{fig:sft_attention_map}, during the reasoning process, the model’s attention to the summary significantly outweighs that allocated to other parts, providing further evidence to support our hypothesis. 

Meanwhile, we believe that historical frames are beneficial for current decision-making. As presented in Tab.~\ref{tbl:history_ablation}, the model achieves a performance of only 30.6 when no historical frames are used, when the number of historical frames is increased to 4 or 5, the performance does not improve further compared to retaining just 2 frames.  Therefore, to balance performance and efficiency, we set the number of historical frames to 2 in all subsequent experiments. Additionally, as shown in Fig.~\ref{fig:sft_attention_map},the attention from actions is more pronounced on the current image compared to historical ones. In contrast, the attention allocated to earlier frames is minimal. This further validates the effectiveness of our data organization strategy.

\begin{wrapfigure}{r}{6.5cm}
  \vspace{-10mm}
  \begin{adjustbox}{max width=\linewidth}
    \includegraphics[width=0.8\textwidth]{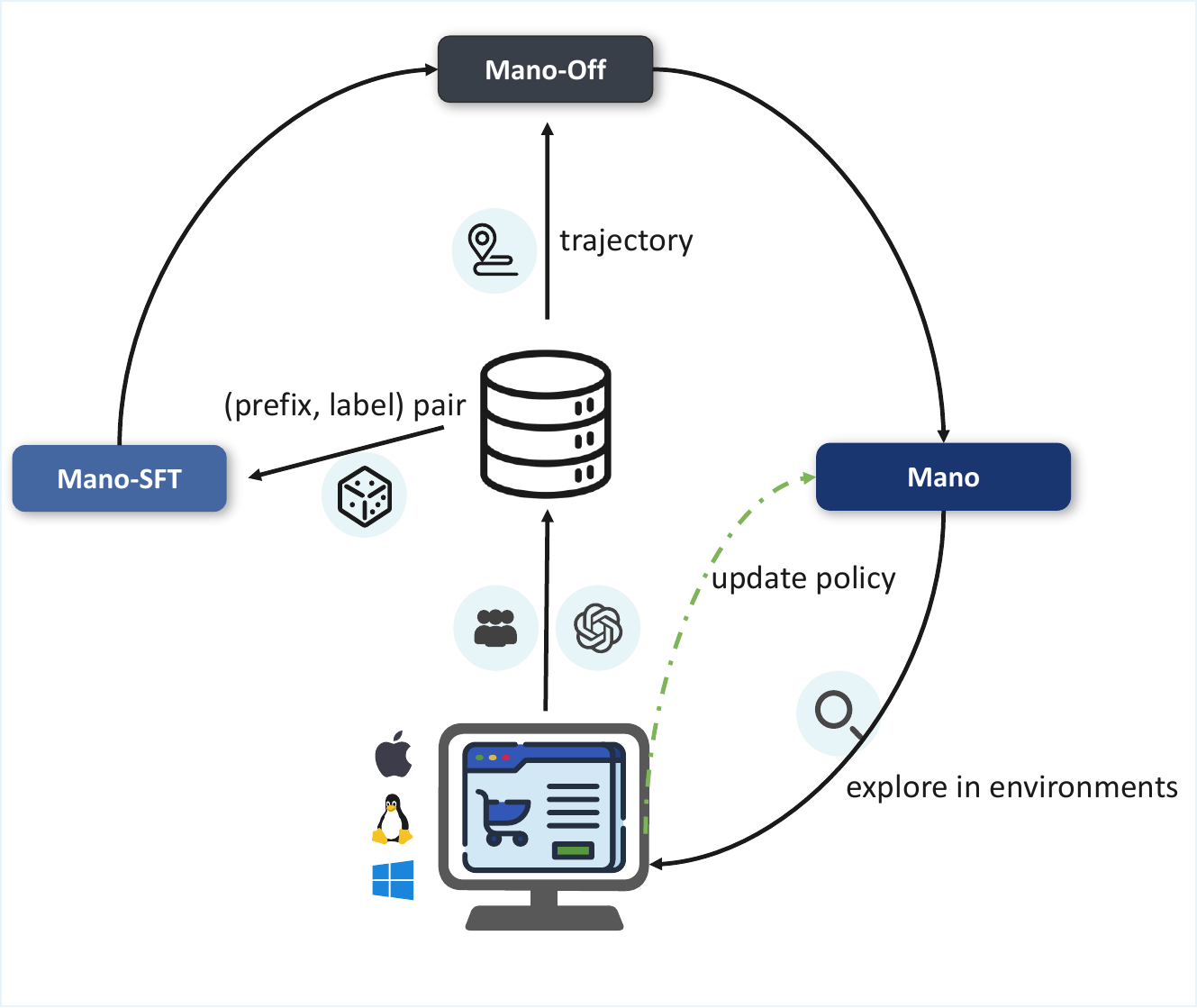}
  \end{adjustbox}
  \vspace{-7mm}
  \caption{Illustration of the data cycling. We sample from the dataset for SFT. Then, complete trajectories are picked to fine-tune the model, which is later used in a simulated environment for RL. During this process, high-value trajectories are retained and are refined through LLM assistance and human correction before being added back to the original dataset.}
  \label{fig:data_loop}
\end{wrapfigure}

\subsection{Closed-loop data cycle}
We have designed a data cycling system, with its framework depicted in Fig.~\ref{fig:data_loop}, which utilizes interactive operation within the GUI web environment to collect decision-centric data. This enhances the model’s capability to proactively perceive and handle real-world stochasticity and non-stationarity during both the SFT and offline RL stages. This approach facilitates continuous self-improvement and iterative learning of the model. Specifically, we retain and reuse trajectories generated during the online RL phase in which the model either executes every step correctly or eventually accomplishes the task despite intermediate errors. To account for the stochastic nature of exploration and to enhance data diversity—--particularly beneficial when training smaller models that may struggle with such samples,we preserve the fully correct trajectories and directly incorporate them into the SFT phase for iterative training. For trajectories that lead to successful outcomes but contain intermediate errors, we refine them through a process where LLM generate initial drafts, which are then reviewed and corrected by human experts before feeding them into the SFT stage. This cyclic process continues until performance gains on our validation set become marginal.

%% file: Styles/sec/data/data_template.tex
\begin{table}[ht]
 \caption{Template for Mano, \textit{prompt} will be replaced with the specific task during training.}
 \label{tbl:template}
{
\setlength{\heavyrulewidth}{1.5pt}
\begin{tabular}{p{\dimexpr\textwidth-2\tabcolsep-2\arrayrulewidth\relax}}
\toprule
You are a GUI agent. You are given a task and your action history, with screenshots. You need to perform the next action to complete the task. \\

\#\# Output Format \\
Thought: ... \\
Action Desp: ... \\
Action: ... \\

\#\# Action Space \\
click(start\_box='<|box\_start|>(x1,y1)<|box\_end|>')\\
type(content='') \\
... \\

\#\# Note \\
- Use English in `Thought` part. \\
- Write a small plan and finally summarize your next action (with its target element) in one sentence in `Action Desp` part. \\

\#\# User Instruction: \\
\textit{prompt} \\

\#\# Action History: \\
step 1: xxx \\
step 2: xxx \\
... \\
step n-2: xxx <image> \\
step n-1: xxx <image> \\
current screenshot is <image> \\
\bottomrule
\end{tabular}
}
\end{table}

%% file: Styles/sec/data/space_def.tex
\begin{table}
\centering
\vspace{-2pt}
\caption{Definition of Action Space for Mano}
\label{tbl:action_def}
\resizebox{\linewidth}{!}{
{
\setlength{\heavyrulewidth}{1.5pt}
\begin{tabular}{lll}
\toprule
\textbf{Action} & \textbf{Params} & \textbf{Description} \\
\midrule
click & Coordinate & Simulates a mouse left-click event. \\
left\_double & Coordinate & Simulates a mouse left-double-click event. \\
right\_single & Coordinate & Simulates a mouse right-click event. \\
right\_double & Coordinate & Simulates a mouse right-double-click event. \\
drag & Start \& target coords & Drags an object to specified location. \\
hotkey & -- & Triggers a hotkey combination. \\
type & Text input & Inputs text and submits with \textbackslash n. \\
scroll & Coord + direction & Simulates mouse wheel scrolling. \\
scroll menu & Coord + direction & Simulates mouse wheel scrolling in specified area. \\
wait & -- & Waits 5s and takes a screenshot to check for changes. \\
call user & -- & Trigger human assistance \\
finish & -- & Finished \\
\shline
\end{tabular}
}
}
\end{table}

%% file: Styles/sec/data/fig/fig_data_engine.tex
\begin{figure}[t]
    \centering
    \includegraphics[width=1.0\linewidth]{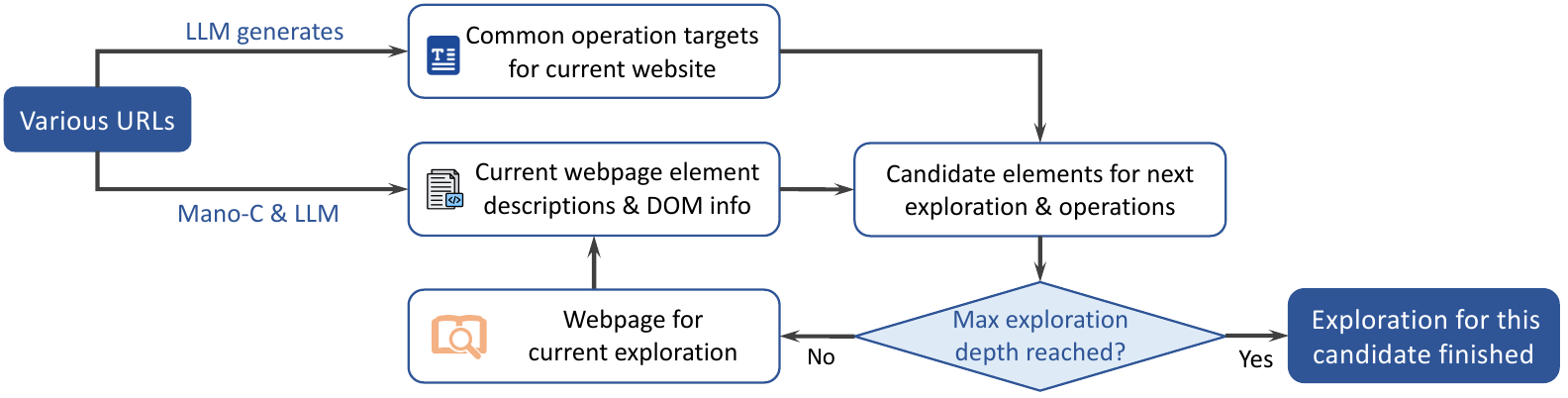}
    \caption{The schematic representation of the automation engine's architecture for collecting web interactions. The process begins with the generation of various URLs by a large language model (LLM). These URLs serve as inputs to identify common operation targets specific to the current website. Subsequently, the system fetches interactive elements on webpages using \textbf{Mano-C}, while also gathering descriptions of current webpage elements along with Document Object Model (DOM) information. This data facilitates the selection of candidate elements for subsequent exploration and operations. The iterative process is governed by a condition that checks whether the maximum exploration depth has been reached. If not, the cycle continues; otherwise, the exploration for the current candidate is deemed complete.}
    \vspace{-4mm}
    \label{fig:data_framework}
\end{figure}

%% file: Styles/sec/data/sft_data_orgs.tex
\begin{table}[ht]
    \vspace{-10pt}
    \centering
    \caption{Data organization during SFT.}
    \tablestyle{10pt}{1.3}
    { 
    \setlength{\heavyrulewidth}{1.5pt}
    \begin{tabular}{lcc}
    \shline
        input & pred \\
        \shline 
         $p_sp_uo_0$& $t_0s_0a_0$\\ 
         $p_sp_uo_0s_0o_1$ &$t_1s_1a_1$\\
         $p_sp_uo_0s_0o_1s_1o_2$ & $t_2s_2a_2$ \\
         $p_sp_us_0o_1s_1o_2s_2o_3$ & $t_3s_3a_3$ \\
         $p_sp_us_0s_1o_2s_2o_3s_3o_4$ & $t_4s_4a_4$ \\
         \vdots & \vdots \\
         $p_sp_us_0s_1s_2\cdots s_{n-5}s_{n-4}o_{n-3}s_{n-3}o_{n-2}s_{n-2}o_{n-1}$ & $t_{n-1}s_{n-1}a_{n-1}$
         \\
    \bottomrule
    \end{tabular}
    }
    \vspace{5pt}
    \label{tbl:sft_data}
    \vspace{-3mm}
\end{table}

%% file: Styles/sec/data/fig/attention_map.tex
\begin{figure}[t]
    \centering
    \includegraphics[width=1.0\linewidth]{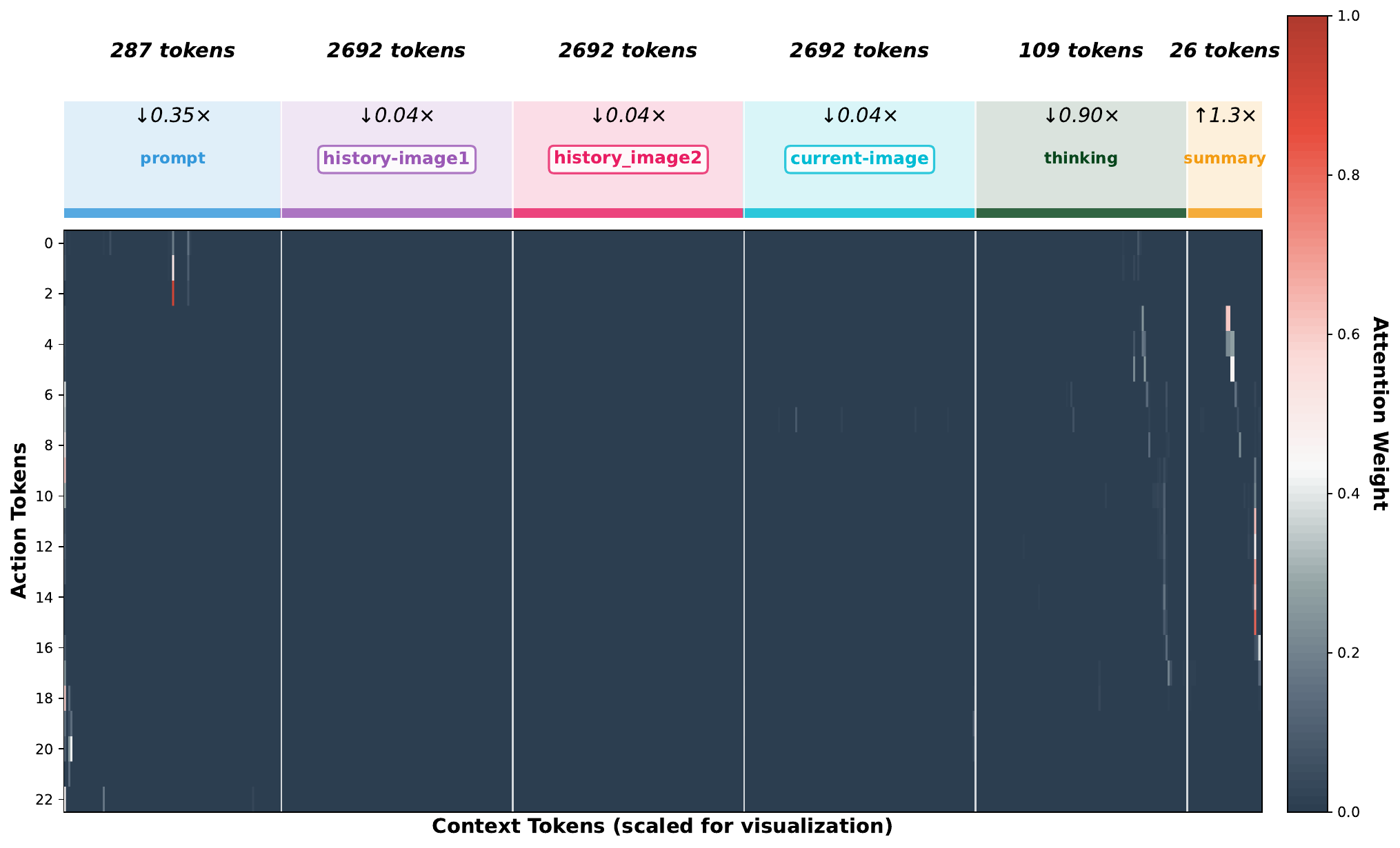}
    \caption{Attention heatmap of \text{action} tokens over prompt, historical/current images, thinking and summary. Note that due to the significantly longer sequence of image tokens compared to the summary and thinking process, the image tokens were compressed via max-pooling-based down-sampling. The vertical axis represents all tokens of the action sequence. Each row corresponds to the attention distribution of one action token over the context. Brighter color indicates higher attention weight.}
    \vspace{-5mm}
    \label{fig:sft_attention_map}
\end{figure}

%% file: Styles/sec/exp/main_exp.tex
We present DeepMiner-Mano-7B, developed through a three-stage training methodology built upon UI-TARS-1.5-7B, encompassing SFT, offline RL, and online RL. For comprehensive evaluation of our model's operational capabilities, we employ two complementary benchmarks: OSWorld-Verified and Mind2Web for web-based operations, which provides extensive coverage across 100+ websites.

\begin{table}[!h]
\Huge
\centering
\caption{Performance comparison across different methods on Mind2Web}\label{table4}
\resizebox{\linewidth}{!}{
{ 
\setlength{\heavyrulewidth}{1.5pt}
\begin{tabular}{lccccccccc}
\toprule
\multirow{2}{*}{Method} & \multicolumn{3}{c}{Cross-Task} & \multicolumn{3}{c}{Cross-Website} & \multicolumn{3}{c}{Cross-Domain} \\
\cline{2-10}
 & Ele.Acc & Op.F1 & Step SR & Ele.Acc & Op.F1 & Step SR & Ele.Acc & Op.F1 & Step SR \\
\midrule
\multicolumn{10}{l}{\textbf{Agent Framework}} \\
GPT-4o~\cite{hurst2024gpt-4o} SeeClick ~\cite{cheng2024seeclick} & 32.1 & - & - & 33.1 & - & - & 33.5 & - & - \\
GPT-4o~\cite{hurst2024gpt-4o} UGround ~\cite{gou2025uground} & 45.7 & - & - & 46.0 & - & - & 46.6 & - & - \\
GPT-4o~\cite{hurst2024gpt-4o} Aria-UI~\cite{yang2025aria} & 57.6 & - & - & 57.7 & - & - & 61.4 & - & - \\
GPT-4V~\cite{2023GPT4Vision} OmniParser~\cite{wan2024omniparser} & 42.4 & 87.6 & 39.4 & 41.0 & 84.8 & 36.5 & 45.5 & 85.7 & 42.0 \\
\midrule
\multicolumn{10}{l}{\textbf{Agent Model}} \\
GPT-4o~\cite{hurst2024gpt-4o} & 5.7 & 77.2 & 4.3 & 5.7 & 79.0 & 3.9 & 5.5 & 86.4 & 4.5 \\
GPT-4~\cite{achiam2023gpt} & 29.6 & - & 20.3 & 20.1 & - & 13.9 & 27.0 & - & 23.7 \\
GPT-3.5(Text-only)~\cite{achiam2023gpt}  & 19.4 & 59.2 & 16.8 & 14.9 & 56.5 & 14.1 & 25.2 & 57.9 & 24.1 \\
GPT-4(Text-only)~\cite{achiam2023gpt} & 40.8 & 63.1 & 32.3 & 30.2 & 61.0 & 27.0 & 35.4 & 61.9 & 29.7 \\
Claude\footnote{Claude refers to Claude-computer-use.}~\cite{Claude3} & 62.7 & 84.7 & 53.5 & 59.5 & 79.6 & 47.7 & 64.5 & 85.4 & 56.4 \\
Aguvis-7B~\cite{xuaguvis} & 64.2 & 89.8 & 60.4 & 60.7 & 88.1 & 54.6 & 60.4 & 89.2 & 56.6 \\
Aguvis-72B~\cite{xuaguvis} & 69.5 & 90.8 & 64.0 & 62.6 & 88.6 & 56.5 & 63.5 & 88.5 & 58.2 \\
CogAgent~\cite{hong2024cogagent} & - & - & 62.3 & - & - & 54 & - & - & 59.4 \\
AutoWebGLM~\cite{lai2024autowebglm} & - & - & 66.4 & - & - & 56.4 & - & - & 55.8 \\
UI-TARS-7B~\cite{qin2025ui} & 73.1 & 92.2 & 67.1 & 68.2 & 90.9 & 61.7 & 66.6 & 90.9 & 60.5 \\
UI-TARS-72B~\cite{qin2025ui} & 74.7 & \textbf{92.5} & 68.6 & 72.4 & 91.2 & 63.5 & 68.9 & \textbf{91.8} & 62.1 \\
\midrule
\rowcolor{gray!20}\textbf{Mano-7B} & \textbf{80.8} & 91.5 & \textbf{73.9} & \textbf{75.7} & \textbf{91.4} & \textbf{68.3} & \textbf{74.3} & 91.5 & \textbf{67.6} \\
\shline
\end{tabular}
}
}
\end{table}

\subsection{Intra-testing}
We introduce 2 benchmarks used for evaluation: multi-modal Mind2Web primarily evaluates the accuracy of web-based operations, covering over 100 websites, 1,000+ operation trajectories, and 7,000+ actions, with 3 testing protocols for comprehensive assessment. OS-Verified comprises 369 evaluation tasks spanning 10 applications, where models execute corresponding tasks in real operating environments, with the final pass rate serving as the evaluation metric. This testing approach more authentically reflects model performance in practical deployment scenarios.

\subsubsection{Browser Use Agent Testing}
The evaluation of multi-modal Mind2Web primarily consists of 3 protocols: cross-task, which measures generalization across tasks within the same environment; cross-website, which evaluates generalization across websites within the same domain; and cross-domain, which assesses generalization across different tasks and environments. We adopt the official evaluation metrics, including element accuracy (Ele.Acc), operation F1 score (Op.F1), and step success rate (Step SR).
Our model is trained on a combination of open-source data, trajectory data automatically collected through Mano-Explorer, and manually annotated operation path data. We employ a three-stage training pipeline consisting of SFT, offline RL, and online RL. At the SFT stage, we mix 10\% open-source data, 70\% automatically collected operation trajectories, and 20\% manually annotated data as the training set, with a learning rate of 1e-5. During the offline RL stage, we select all samples with grounding errors and samples with operational step errors during the planning process from the SFT stage as the training set, employing GRPO for training with a group size of 8. At online RL stage, we utilize trajectories from interactions in virtual environments using both automatically collected and manually annotated samples as the final training set, thereby adapting to variations in real virtualized environments.
Our trained DeepMiner-Mano-7B is compared against SOTA methods based on both agent frameworks and agent models on the Mind2Web test set. The results demonstrate that our method achieves significant improvements in operation accuracy compared to SOTA approaches, attributed to the incorporation of extensive online data augmentation for coordinate element localization training and the introduction of RL to enhance localization precision. The accuracy of operation types remains comparable to SOTA methods, while the overall success rate shows substantial improvement. Detailed results are presented in Tab.~\ref{table4}.

\begin{table}[h]
\centering
\caption{Performance on OSWorld (Foundation E2E GUI \& Specialized model)}\label{table5}
\resizebox{\linewidth}{!}{
{ 
            \setlength{\heavyrulewidth}{1.5pt}
\begin{tabular}{llcc}
\toprule
Method & Approach \& Details & Success Rate (Avg±Std)\\
\midrule
opencua-qwen2-7b & Type: Specialized model, Max Steps: 100, Runs: 1 & 23.1 \\
UI-TARS-7B & Type: Specialized model, Max Steps: 100, Runs: 2 & 27.4±2.2 \\
uitars-72b-dpo & Type: Specialized model, Max Steps: 100, Runs: 1 & 27.1 \\
TianXi-Action-7B & Type: Specialized model, Max Steps: 50, Runs: 2 & 29.8±0.6 \\
computer-use-preview & Type: Specialized model, Max Steps: 50, Runs: 1 & 31.3 \\
GUI-Owl-7B & Type: Specialized model, Max Steps: 15, Runs: 1 & 32.1 \\
opencua-32b & Type: Specialized model, Max Steps: 100, Runs: 3 & 34.8±0.8 \\
\midrule
\rowcolor{gray!20} \textbf{Mano-7B} & Type: Specialized model, Max Steps: 100, Runs: 2 & \textbf{41.6±0.7} \\
\rowcolor{gray!20} \textbf{Mano-72B} & Type: Specialized model, Max Steps: 100, Runs: 1 & \textbf{53.8} \\
\midrule
\end{tabular}
}
}
\vspace{-5mm}
\end{table}

\subsubsection{Computer Use Agent Testing}
We employ OSWorld-Verified as the benchmark for evaluating model performance on CUA end-to-end tasks. The evaluation executes corresponding tasks in batches within Ubuntu virtual environments, with the final evaluation metric being the average score. This metric represents the mean completion score across all tasks in the evaluation set, where each task typically receives 100 points for successful verification, 0 points for failure, with certain tasks allowing intermediate scores.
During the CUA training phase, we performed action space alignment, data organization restructuring, and reasoning component adjustments on the open-source trajectory data provided by OpenCUA~\cite{wang2025opencuaopenfoundationscomputeruse}, which constitutes 30\% of the overall operation trajectory data. Additionally, we incorporate automatically collected trajectory data, accounting for 30\% of the total training data, and manually collected computer operation trajectories, comprising 40\% of the total training data. We also include grounding data corresponding to the interfaces in the trajectories. In the training phase, we adopt the same training approach and strategies as BUA to obtain the final model. We benchmarked our approach against SOTA models from the Foundation E2E GUI \& Specialized model track on the OSWorld-Verified leaderboard, with comparative results presented in Tab.~\ref{table5}. 
Notably, the newly introduced Mano-72B achieves success rate of 53.8, surpassing the Mano-7B model’s 41.6, further demonstrating that our method consistently benefits from larger model capacities.

\begin{table}[h]
    \centering
    
    \begin{minipage}[c]{0.43\columnwidth}
        \centering
        \caption{Scaling of historical images}
        \label{tbl:history_ablation}
        { 
            \setlength{\heavyrulewidth}{1.5pt} 
            \begin{tabular}{cc}
            \toprule 
            \shortstack{Number of historical images} & \shortstack{Avg Score} \\
            \midrule
             0 & 29.6 \\
             1 & 31.5 \\
             2 & 32.7 \\
             3 & 32.6 \\
             4 & 32.7 \\
            \bottomrule
            \end{tabular}
        } 
    \end{minipage}
    \hfill 
    \begin{minipage}[c]{0.55\columnwidth}
        \centering
        \caption{SFT Training Ablation}
        \label{tbl:sft_ablation}
        { 
            \setlength{\heavyrulewidth}{1.5pt} 
        \begin{tabular}{lc}
        \toprule
         Method & \shortstack{Avg Score} \\
        \midrule
         Baseline & 25.1\\
         SFT (UITARS-1.5-7B) & 29.9 \\
         \rowcolor{gray!20} SFT (DeepMiner-Mano-7B) & 32.7\\
        \bottomrule
        \end{tabular}
        }
    \end{minipage}
    \vspace{-5mm}
\end{table}

\begin{table}[h]
    \vspace{-4mm}
    \centering
    \caption{Performance comparison across different training phases on OSWorld-Verified.}
    {
    \setlength{\heavyrulewidth}{1.5pt}
    \begin{tabular}{llc}
    \shline
        stage & Method &Avg score \\
        \shline 
         --&Baseline & 25.1\\ 
         \midrule
         Stage 1&SFT & 32.7\\
         Stage 2&Offline RL & 33.7\\
         \rowcolor{gray!20} Stage 3&Online RL & 41.6 \\
    \bottomrule
    \end{tabular}
    }
    \vspace{5pt}
    \label{tbl:stage_ablation}
    \vspace{-12pt}
\end{table}

\subsection{Ablation Study}
We conducted ablation experiments on the following modules. (i). For the number of historical images used in operation history modeling, we evaluated models with 0-4 historical images as context on OSWorld-Verified, finding that 2 historical images yielded optimal performance, as shown in Tab.~\ref{tbl:history_ablation}. (ii).Under identical training data conditions, we performed SFT using two distinct approaches: UITARS's multi-turn dialogue framework for organizing historical operations and screenshots, and our proposed method. Comparative analysis reveals that our approach—which summarizes historical steps, reduces the quantity of historical images, and integrates reflection on historical data within the thinking module—yields a 2.8 percentage point performance gain. Detailed comparative results are presented in Tab.~\ref{tbl:sft_ablation}. (iii). We assessed performance improvements on OSWorld-Verified after different learning stages of the model. Results indicate that benefiting from CUA training data expansion, adjusted historical operation modeling approaches, the SFT stage alone achieved a significant improvement of 7.6 points. The offline RL stage showed only a 1-point improvement, primarily because the performance on SFT-stage training data was approaching saturation, with improvements limited to particularly challenging element localization cases. To address this limitation, we sampled more diverse data through interactions with online environments. Experimental results demonstrate that online RL contributed a substantial improvement of 7.9 points. Detailed comparative results are presented in Tab.~\ref{tbl:stage_ablation}.

\vspace{-2mm}
\begin{figure}[h]
    \centering
    \includegraphics[width=1.00\linewidth]{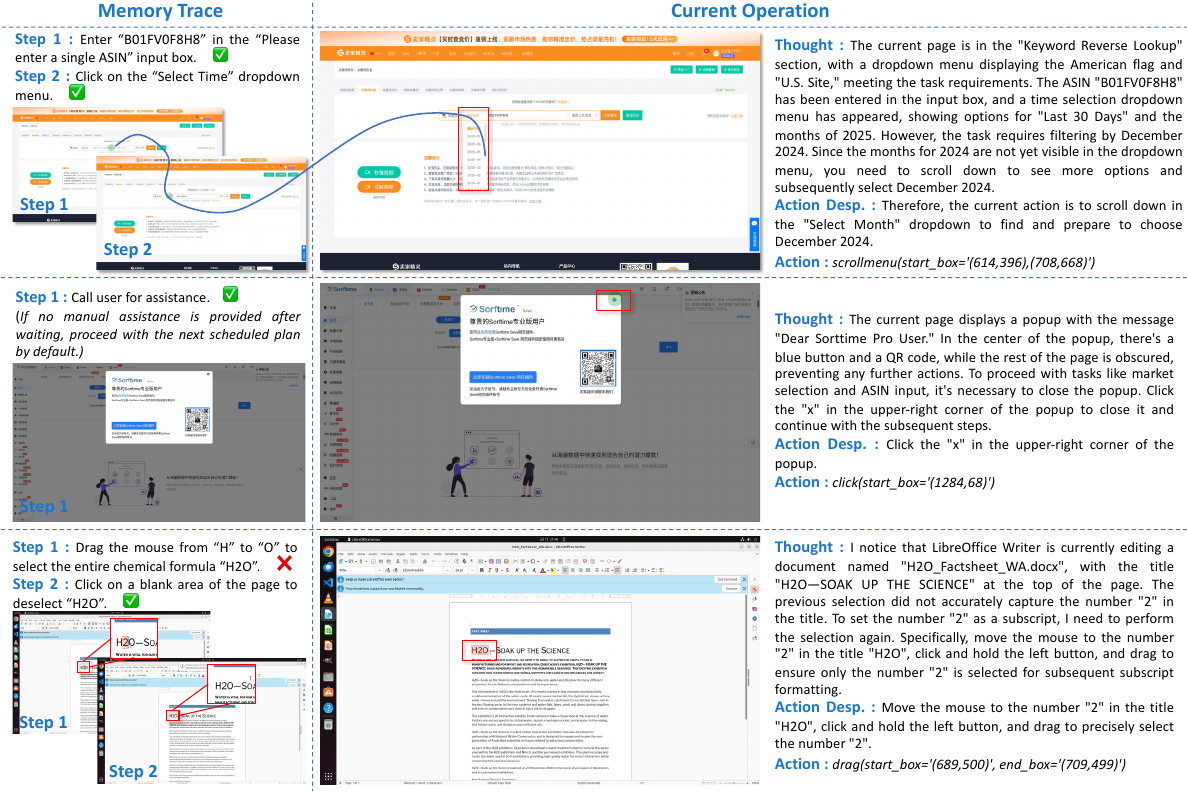}
    \caption{Illustrative examples of Mano’s reasoning and execution process across 3 distinct scenarios. Each row corresponds to one task instance: the left column records the historical trace of executed actions with verification marks, while the right column displays the current reasoning state, including generated \emph{Thought}, action description, and executable function. Row 1 demonstrates environment extension by scrolling a dropdown to reveal a hidden option; Row 2 shows exception handling and generalization when facing an unexpected popup; Row 3 highlights self-analysis and correction after an erroneous selection. Together, these cases illustrate Mano’s robustness, adaptability, and error-aware reasoning in complex interactive environments.}
    \vspace{-4mm}
    \label{fig:analysis_examples}
\end{figure}

\subsection{Analysis and Visualization}
Figure~\ref{fig:analysis_examples} presents 3 representative examples that illustrate the reasoning dynamics, error handling, and generalization ability of the proposed Mano in real-world scenarios. Each row in the figure corresponds to a complete task trace, where the left column records the historical action sequence (denoted as \emph{Memory Trace}), and the right column documents the current decision-making state (\emph{Current Operation}). In the historical trace, every step is annotated with the natural-language description of the executed action (Action Desp or summary), followed by the verification results of Mano-verify, which automatically checks the correctness of execution (success and failure). On the right-hand side, the current operation displays the agent’s generated \emph{Thought}, the translated action description, and the executable action function along with explicit parameters such as coordinate locations. By jointly examining these two perspectives, we can directly observe how Mano integrates memory-based reasoning, current-state perception, and self-correction to accomplish complex tasks.

\subsubsection{Case 1: Controlled Completion via Menu Expansion}

The first row illustrates a standard case of interaction with a structured web interface. The task requires the system to input a specific ASIN code into the designated text field, open a time-selection dropdown menu, and subsequently choose a temporal option that does not initially exist in the interface (\texttt{2024-12}). 
From the historical trace, we observe the sequential steps of text input and dropdown activation, both successfully executed and validated by Mano-verify. At the current stage, however, the required temporal option is missing in the visible range. The agent thus reasons that scrolling is necessary to reveal additional menu items. The generated action specifies a direct coordinate-based scrolling command ( $scrollmenu(start\_box=(614,396), end\_box=(708,668))$  ), which ensures deterministic manipulation of the dropdown interface. This example highlights Mano’s ability to augment incomplete environments by combining domain-specific priors (the existence of the month “2024-12”) with low-level executable controls. The precise mapping from high-level intention (“select 2024-12”) to low-level actionable coordinates exemplifies how our design enables robust grounding across heterogeneous UI structures.

\subsubsection{Case 2: Exception Handling and Generalization}

The second row demonstrates a scenario that falls completely outside the training distribution: the unexpected appearance of a modal popup that obscures the interface. Such irregular conditions are common in real-world software environments, where unanticipated dialogues, update notifications, or advertisements may interrupt task execution. In this case, the initial system response is to invoke ( $ call\_user()$ ), an action representing explicit escalation for human intervention. This decision reflects Mano’s design principle of fail-safe delegation, where safety and accuracy take precedence over forced autonomous operation. Importantly, the figure also shows that after a certain waiting period without manual assistance, the agent autonomously reevaluates the environment and proposes to close the popup by clicking the “x” in the upper-right corner. This adaptive behavior illustrates Mano’s generalization capacity: despite never encountering such a configuration during training, it successfully extrapolates an appropriate action by leveraging its perception-action alignment and failure-recovery strategies. The coexistence of human-in-the-loop fallbacks and autonomous recovery mechanisms demonstrates the system’s robustness to environmental uncertainty.

\subsubsection{Case 3: Self-Analysis and Error Correction}

The third row presents an instructive case of error correction within a document-editing environment (LibreOffice Writer). The task involves selecting the subscript “2” in the chemical formula “H\textsubscript{2}O”. Initially, the agent incorrectly selects the entire token “H2O”, as confirmed by Mano-verify with a failure mark. Crucially, rather than persisting in the erroneous behavior, Mano engages in explicit self-analysis, recognizing that the prior selection exceeded the intended scope. In its subsequent reasoning, the agent refines the action to precisely drag-select the single character “2” by adjusting the bounding box coordinates. This correction not only achieves the intended goal but also illustrates the model’s capacity for reflective reasoning: by integrating feedback signals (from the verification module) with its internal plan, it can update its policy on the fly. This form of error-aware self-adjustment is particularly valuable in dynamic editing tasks, where fine-grained precision is required, and even minor deviations can compromise the semantic correctness of results.

\subsubsection{Discussion}

Collectively, the 3 cases reveal distinct yet complementary dimensions of Mano’s operational reliability. Case 1 emphasizes environment extension through low-level grounding, enabling the system to interact with incomplete or hidden UI elements. Case 2 illustrates resilience to unexpected disturbances, where the agent can flexibly transition between human-assisted and autonomous modes. Case 3 demonstrates reflective correction, showing that the system can internalize mistakes and rectify them within a single execution trajectory. Taken together, these observations highlight the effectiveness of our design philosophy: integrating structured memory, explicit reasoning chains, and feedback-driven correction mechanisms to achieve robust, generalizable, and self-adaptive action execution.

These examples also underscore the methodological significance of joint visualization. By juxtaposing \emph{Memory Trace} and \emph{Current Operation}, we make the internal reasoning processes of the agent transparent to both developers and users. This transparency not only facilitates debugging and evaluation but also provides a clear foundation for trust and accountability in practical deployments.

%% file: Styles/sec/conclusion/con.tex
Mano represents a paradigm shift in the development of robust, general-purpose Graphical User Interface (GUI) agents. By seamlessly integrating a state-of-the-art multimodal foundation model with a meticulously structured three-stage training pipeline, Mano achieves unprecedented performance in GUI interaction tasks. The training pipeline, comprising SFT, offline RL, and online RL, is augmented by carefully crafted reward designs and a novel simulated environment. This comprehensive approach enables Mano to achieve strong alignment with GUI-specific domains, enhance multistep reasoning capabilities, and significantly improve adaptability to dynamic interfaces.

Rigorous empirical evaluations conducted on established benchmarks, including Mind2Web and OSWorld, demonstrate Mano's superiority in multiple key performance metrics. In particular, Mano sets new state-of-the-art standards in success rate, element precision, and stepwise completion. Extensive ablation studies corroborate the efficacy of critical design choices, such as the implementation of online RL, the utilization of historical context, the application of attention constraints and the incorporation of a closed-loop data cycle. Each of these components contributes substantially to the agent's overall performance, underscoring the synergistic nature of Mano's architecture.

The remarkable results achieved by Mano emphasize the critical importance of domain-specific data generation, iterative training through reinforcement learning, and holistic reward design to overcome the inherent limitations of conventional vision language models in GUI interaction tasks. Beyond providing a scalable and efficient framework for GUI automation, Mano offers valuable insights into the symbiotic relationship between imitation learning and reinforcement learning in the context of embodied AI systems.

Future research directions for the Mano project are multifaceted and promising. We plan to elucidate the data acquisition capabilities of Mano-parking in greater detail, proposing a novel benchmark that positions data acquisition as the primary objective. Additionally, we aim to provide comprehensive insights into the training methodology and the reasoning process integration of Mano-verify. Furthermore, we intend to expand Mano's functionality by introducing Mano-cipher and its automated login capabilities, thereby enhancing its applicability in real-world environments and enhancing the on-device deployment ability based on model compression techniques~\cite{yu2025mquant}.